# Probing Charging and Localization in the Quantum Hall Regime by Graphene *pnp* Junctions

Jairo Velasco Jr., Gang Liu, Lei Jing, Philip Kratz, Wenzhong Bao, Marc Bockrath, Chun Ning Lau[1*]

Department of Physics and Astronomy, University of California, Riverside, CA 92521

* Email: lau@physics.ucr.edu

**Using high quality graphene *pnp* junctions, we observe prominent conductance fluctuations on transitions between quantum Hall (QH) plateaus as the top gate voltage $V_{tg}$ is varied. In the $V_{tg}$-$B$ plane, the fluctuations form crisscrossing lines that are parallel to those of the adjacent plateaus, with different temperature dependences for the conductance peaks and valleys. These fluctuations arise from Coulomb-induced charging of electron- or hole-doped localized states when the device bulk is delocalized, underscoring the importance of electronic interactions in graphene in the QH regime.**

Graphene's unique electrical[1, 2], thermal[3] and mechanical[4, 5] properties have made it a topic of intense focus. Electronic interactions in graphene have been predicted to play important roles in its electronic properties, because of the enhanced Coulomb repulsion and reduced screening in this strictly two-dimensional (2D) system. In particular, interaction effects are predicted to have profound effects on graphene in the quantum Hall (QH) regime[6, 7]. The recent observation of fractional quantum Hall effect in high mobility suspended graphene devices[8, 9] has confirmed the importance of electronic interaction in graphene in the QH regime. Nevertheless, most of the experimental results in graphene in the integer QH regimes can be satisfactorily interpreted by single particle pictures[10, 11], and the evidence for interaction effects has remained surprisingly elusive.

In this Letter we present conductance measurement of high quality graphene *pnp* junctions[12-17] with suspended top gates in the QH regime. In magnetic fields $B>3T$, the device display well-developed conductance plateaus at uniform charge density $n$, with up to 15 discernible plateaus. Interestingly, we observe reproducible conductance fluctuations between quantum Hall plateaus when the charge density of the top-gated region $n_{tg}$ is modulated. In the $n_{tg}$-$B$ plane, slopes of these fluctuations are parallel to those of the center of the adjacent plateaus with integer filling factors. We attribute such QH fluctuations to Coulomb-induced charging of localized states in the device[18-20], which are estimated to be 80-130nm in diameter. Notably, for the zeroth Landau level (LL), both positive and negative slopes in the $n_{tg}$-$B$ plane are observed, indicating the evolution of both hole- and electron-doped states. Moreover, with increasing temperature, the conductance peaks of the fluctuations remain constant, while the conductance valleys increase, in agreement with the behavior of a Coulomb-blockaded quantum dot. Our results reflect the importance of electronic interaction in QH regime in graphene, and demonstrate that electrical transport in *pnp* junctions uniquely reveals both local and global properties of the devices.

The substrate consists of a 300-nm $SiO_2$ layer grown over degenerately doped silicon, which serves as a global back gate. Graphene sheets are deposited onto the substrate by the standard micro-mechanical exfoliation method[10, 21], and identified by color contrast in an optical microscope and Raman spectroscopy[22]. The electrodes, which consist of 10 nm of Ti and 80 nm of Al, are deposited on graphene using standard electron beam lithography. A multi-level lithography process[23, 24] is used to fabricate a suspended top gate that straddles the center portion of graphene (Fig. 1a-b). Since no dielectric material directly contacts the single atomic layer, these graphene devices are exceedingly clean[24]. The air-gap between the top gate and graphene also allows surface absorbates, including those directly under the top gate, to be removed via annealing[25]. As a demonstration, we measure the two-terminal conductance *G* of a graphene device as a function of applied back gate voltage $V_{bg}$ both before and after annealing in vacuum at 120 ºC (Fig. 1c). The $G(V_{bg})$ characteristics is considerably improved upon annealing, with a sharper Dirac point that is closer to zero, indicating the removal of undesirable absorbates and resist residue.

The graphene devices are measured at 260mK using standard lock-in techniques. Similar data are observed in 5 devices. Here we focus on data from a single device, with source-drain separation *L*=1.4 µm; width of the graphene sheet is 1.8 µm at the source and 1.4 µm at the drain electrodes, respectively. The top gate is suspended at a distance *t*~100 nm above the substrate, covering a 400-nm-long segment of the device.

At high magnetic fields, graphene's relativistic band structure gives rise to Landau levels (LLs) with energies $E_N = \text{sgn}(N)\sqrt{2e\hbar v_F^2 |N| B}$ [2]; between the LLs, the device conductance is quantized at $4(N+\frac{1}{2})\frac{e^2}{h}$. Here $N$ is an integer denoting the LL index, $e$ the electron charge, $h$ Planck's constant, and $v_F \sim 10^6$ m/s the Fermi velocity of the charges. The upper panel of Fig. 1d plots $G$ in units of $e^2/h$ vs. $V_{bg}$ at $B$=8T with the top gate disconnected, and clear plateaus at $v$=2,6,10, 14 and 18 with almost perfectly quantized conductances are observed, where $v = \frac{nh}{Be}$ is the filling factor and $n$ is the electrostatically-induced charged density in graphene. The standard "fan diagram" of QH plateaus is shown in the lower panel of Fig. 1d, for $B$ ranges from 2 to 8T and $V_{bg}$ from -5 to 50V. A total of 15 QH plateaus are discernible (though conductance for plateaus for $v$>20 are not exactly quantized), and the $2e^2/h$ plateau is quantized at magnetic field as low as 2.5T. All these features underscore the high quality of the device.

From the fan diagram, the plateaus' centers have slopes in the $V_{bg}$-$B$ plane given by

$$b_{bg}=ve/h\alpha_{bg} \qquad (1)$$

where $\alpha_{bg}=n/V_{bg}$ is the coupling efficiencies for the back and top gates, respectively, and $v=4(N+\frac{1}{2})$=2, 6, 10… In Fig. 1d, for the $N$=0, 1 and 2 plateaus, the slopes are measured to be $b_{bg}$=0.74, 2.0, and 3.4, respectively, yielding $\alpha_{bg} \sim 7.5\times10^{10}$ cm$^{-2}$ V$^{-1}$, in good agreement with that estimated from geometric considerations.

By applying voltages to both top and back gates, we modulate the induced charge density in the top gated and non-top-gated regions, thus creating *pnp* junctions with *in situ* tunability and dopant levels. In the QH regime, the non-uniform charge density gives rise to regions of different filling factors, and, for bi-polar (i.e. *pnp* or *npn*) junctions, counter-propagating edge states. Consequently, the device conductance exhibit plateaus at fractional values of $e^2/h$ that arise from the mixing of edge states at the interfaces. Assuming full edge stage equilibration, the device conductance is given by[13]

$$G = e^2/h |\nu_2| \qquad \text{if } \nu_1\nu_2>0 \text{ and } |\nu_1| \geq |\nu_2|, \qquad (2a)$$

$$G = \frac{e^2}{h}\left(\frac{1}{|\nu_1|} \mp \frac{1}{|\nu_2|} + \frac{1}{|\nu_1|}\right)^{-1} \quad \begin{cases} -: \text{ if } \nu_1\nu_2 > 0 \text{ and } |\nu_2|>|\nu_1| \\ +: \text{ if } \nu_1\nu_2 L < 0 \end{cases} \qquad (2b)$$

where $\nu_1$ and $\nu_2$ are the filling factors in the areas outside and within the top-gated regions, respectively. As shown in Fig. 2a, a typical QH map $G(V_{bg}, V_{tg})$ at $B=8$T appears as a plaque of adjoined parallelograms with different colors, representing QH plateaus with different combination of $\nu_1$ and $\nu_2$. The slope of the diagonal lines, which is measured to be ~0.8, yields the ratio of the coupling efficiencies between the two gates. This agrees with that estimated from a simple model of parallel plate capacitance, given by $\eta=d/(t\varepsilon_{siO})$ ~0.77, where $d=300$ nm and $\varepsilon_{siO}=3.9$ are the thickness and dielectric constant of the SiO$_2$ layer. Thus the top gate efficiency $\alpha_{tg}=n/V_{tg}$ is estimated to be $6\times10^{10}$ cm$^{-2}$ V$^{-1}$. Three line traces $G(V_{tg})$ at $\nu_1=2$, 6 and 10 are displayed in the lower panel of Fig. 2a. The measured conductance values of the plateaus are in excellent agreement with that obtained from Eq. (2). Though similar data have been reported before[12, 13, 26, 27], we would like to emphasize that this is the first data set to date that demonstrates fully developed and properly quantized plateaus up to the $N=2$ LL in graphene *npn* and *nn'n* junctions, again underscoring the high junction quality.

Interestingly, though the transitions between QH plateaus appear to be smooth in Fig. 2a, more complicated structures develop upon high resolution $G(V_{bg}, V_{tg})$ scans. Fig. 2b presents data over a smaller range of that in Fig. 2a, but taken with 10 times higher gate voltage resolution. Instead of smooth steps, the transitions between different QH plateaus now display pronounced fluctuations. A line trace of the transition between $G=2e^2/h$ to $6e^2/h$ plateaus is shown in the lower panel (red curve). To identify the movement of such fluctuations more clearly, we plot $dG/dV_{tg}$ in the middle panel of Fig. 2b. Evidently, these fluctuations are observed on all the transitions between different plateaus. Moreover, even though similar conductance fluctuations can be observed in high resolution $G(V_{bg})$ plots when the device has uniform charge density (*i.e.* with the top gate disconnected), the amplitudes are significantly smaller, indicating the local nature of these fluctuations.

We note that with medium resolution scans, these fluctuations appear as kinks or small plateaus on the transition, as shown by the blue curve in Fig. 2b, and may resemble that of spin- or valley-resolved LLs. To elucidate the origin of these fluctuations, we study their evolution with $V_{tg}$ and $B$ at constant $V_{bg}$. Fig. 3a plots the differentiated conductance $dG/dV_{tg}$ as functions of $V_{tg}$ and $B$ at $V_{bg}$=2.5V, or equivalently, at $\nu_1$=2. At $B$=8T, the conductance changes from $0.67e^2/h$ at $V_{tg}$=-13V (or $\nu_2 \sim$ -1.7) to $2e^2/h$ at $V_{tg}$=-7V (or $\nu_2 \sim$0.2). Fig. 3b plots another data set $dG/dV_{tg}(V_{tg}, B)$, taken at $V_{bg}$=14V, or $\nu_1$=6. At $B$=8T, the conductance increases from $2e^2/h$ at $V_{tg}$=-20V (or $\nu_2 \sim$0.5) to $2e^2/h$ at $V_{tg}$=-7V (or $\nu_2 \sim$4.6). In both graphs, two pronounced features become evident: "smooth" sections that increase in area with increasing magnetic field, separated by strips of "rough areas" that stay approximately constant in size. In the smooth regions, $dG/dV_{tg}$=0, so they correspond to QH plateaus in the locally gated area; as the energetic spacing between LLs increases with magnetic fields, the width of a given smooth region grows accordingly. The red dotted lines plots the slopes of the plateaus in the $V_{tg}$-$B$ plane, given by $b_{bg}/\eta$, which are calculated using values of $b_{bg}$ measured from Fig. 1d, and $\eta$ =0.8 from Fig. 2a.

We now focus on the "rough" areas separating the QH plateaus in Fig. 3. They consist of crisscrossing bright and dark lines, corresponding to extrema values in conductance fluctuations between QH transitions. Noticeably, trajectories of these ridges are parallel to the dotted lines, or to the adjacent QH plateaus. For instance, in Fig. 3b, the rough area separating the 2 $e^2/h$ and 6 $e^2/h$ plateaus consist of lines with the same slopes as the $\nu_2$=2 and $\nu_2$=6 plateaus; in Fig. 3a, the lines are parallel to the $\nu_2$=-2 and 2 plateaus, respectively. In the low-$B$ (lower left) corner of Fig.

3b, several short lines parallel to the $v_2$=2 plateau are visible, but abruptly stop as the device conductance become quantized at higher fields.

Such conductance fluctuations between QH transitions have been observed in MOSFET devices[18], though only with positive slopes. In the single-particle picture, conductance fluctuations may arise from resonant tunneling between the edges states on opposite sides of the device through states located in the bulk[28]. However, trajectories of these fluctuations are expected to be parallel to that of half-integer values of filling factors[18]. Thus, we attribute these fluctuations to charging and localization induced by electronic interactions. Assuming a general disorder potential induced by, *e.g.* impurities on the substrate or graphene, the electron density in graphene develop local valleys and hills. At high magnetic fields, the edge channels in a 2D electron system form compressible and incompressible strips, with contours that generally follow the local potential landscapes. Thus, when an incompressible (insulating) strip completely surrounds a compressible (metallic) region, this effectively creates a quantum dot with quantized charges. The conductance fluctuations then arise from charging of and transport across one or multiple such quantum dots, and should only depend on the geometry of the dots. Since the filling factor $v$ within an incompressible strip takes on integer values, which only depends on the ratio $n/B$, the size of the dot (and hence the fluctuations) should remain constant for the same $n/B$ value, as observed experimentally. We note that in contrast to MOSFET devices[18], fluctuations with both positive and negative slopes in the $V_{tg}$-$B$ plane are observed, indicating the presence of both electron- and hole- doped incompressible strips at the $N$=0 LL, a manifestation of graphene's unique band structure.

Our data shown in Fig. 3b, *i.e.,* the parallel lines with slope given by Eq. (1) for $v$=-2, 2, 6…, bear striking resemblance with that of local inverse electronic compressibility of graphene[19] and GaAs/AlGaAs[20] devices using scanned single electron transistors (SET), albeit with one important difference: in the SET measurements, the line appear on the QH plateaus, at which the bulk of the device consists of incompressible states; in our experiment, the lines appear at the transitions, or equivalently, at the center of the Laudau Levels, when states in the bulk of the graphene are metallic and delocalized.

More supporting evidence is given by the quasi-periodic nature of some of the conductance fluctuations, with a typical separation in $V_{tg}$ of $\Delta V_{tg}$~0.1 – 0.35 V. In the Coulomb blockade scenario, this represents the charging of a compressible quantum dot that dominates transport. Addition of one electron requires $\Delta V_{tg} = e/(\varepsilon_0 A/t)$, where $\varepsilon_0$ is the permittivity of vacuum and $A$ is the area of the metallic region. This yields a typical metallic area with a diameter of ~80 – 135nm, consistent with SET measurements[19, 29]. Moreover, when a LL is almost completely filled, an incompressible strip percolates through the metallic regions, marking the onset of QH conductance plateaus. This corresponds to the short parallel lines that abruptly terminate just before the onset of QH conductance plateaus, as seen in the lower left corner of Fig. 3b.

Finally, we study the temperature dependence of these fluctuations. We focus on the transition between $v_2$=2 and $v_2$=6 plateaus, whereas the non-top-gated regions are kept at $v_1$=6. The data are taken at $B$=8T and 10 different temperatures $T$ between 0.26K and 3.8K(Fig. 4a). In particular, we follow the evolution of the group of fluctuations near the center of the transition, as marked by the arrows on Fig. 4a. Interestingly, with increasing $T$, the conductance of the peak of these fluctuations $G_p$ remains approximately constant, while that of the dips $G_d$ increases (Fig. 4b). This is exactly that expected for a Coulomb-blockaded classical quantum dot[30]: the dot's maximum conductance is temperature independent, as it is simply determined by the ohmic addition of the barrier resistances; the minimum conductance of the dot, however, is thermally activated. Thus, our data suggest that a single quantum dot dominates the conductance fluctuations. Modeling the conductance between the QH transitions by that of a single dot in parallel with other metallic paths, we write $G_d$ as[30]

$$G_d(T) = \frac{A/k_B T}{\sinh(\delta/k_B T)} + C \qquad (3)$$

where $\delta$ is the energy detuning from the resonant level, and $k_B$ is the Boltzmann constant. Using $A=123k_B$, $\delta=4.0k_B$, and $C=98$ μS, we plot the resultant curve as the green line in Fig. 4b, which is in excellent agreement with the data.

We thank Gil Refael and Shan-Wen Tsai for helpful discussions, and Peng Wei for assistance with LABVIEW software. The work is supported in part by NSF CAREER DMR/0748910, NSF/ECCS 0926056, ONR N00014-09-1-0724 and UCOP.

**Reference**

1. K. S. Novoselov *et al.*, Science **306**, 666 (2004).

2. A. H. Castro Neto *et al.*, Rev. Mod. Phys. **81**, 109 (2009).

3. A. A. Balandin *et al.*, Nano Lett. **8**, 902 (2008).

4. C. Lee *et al.*, Science **321**, 385 (2008).

5. W. Z. Bao *et al.*, Nature Nanotechnol. **4**, 562 (2009).

6. Y. Zhang *et al.*, Phys. Rev. Lett. **96**, 136806 (2006).

7. K. Nomura and A. H. MacDonald, Phys. Rev. Lett. **96**, 256602 (2006).

8. X. Du *et al.*, Nature **462**, 192 (2009).

9. K. I. Bolotin *et al.*, Nature **462**, 196 (2009).

10. K. S. Novoselov *et al.*, Nature **438**, 197 (2005).

11. Y. B. Zhang *et al.*, Nature **438**, 201 (2005).

12. J. R. Williams, L. DiCarlo, and C. M. Marcus, Science **317**, 638 (2007).

13. B. Ozyilmaz *et al.*, Phys. Rev. Lett. **99**, 166804 (2007).

14. B. Huard *et al.*, Phys. Rev. Lett. **98**, 236803 (2007).

15. R. V. Gorbachev *et al.*, Nano Lett. **8**, 1995 (2008).

16. D. A. Abanin and L. S. Levitov, Science **317**, 641 (2007).

17. V. V. Cheianov and V. I. Fal'ko, Phys. Rev. B **74**, 041403 (2006).

18. D. H. Cobden, C. H. W. Barnes, and C. J. B. Ford, Phys. Rev. Lett. **82**, 4695 (1999).

19. J. Martin *et al.*, Nature Phys. **5**, 669 (2009).

20. S. Ilani *et al.*, Nature **427**, 328 (2004).

21. F. Miao *et al.*, Science **317**, 1530 (2007).

22. A. C. Ferrari *et al.*, Phys. Rev. Lett. **97**, 187401 (2006).

23. G. Liu *et al.*, Appli. Phys. Lett. **92**, 203103 (2008).

24. J. Velasco *et al.*, New J. Phys. **11**, 095008 (2009).

25. J. H. Chen *et al.*, Nature Phys. **4**, 377 (2008).

26. D. K. Ki and H. J. Lee, Phys. Rev. B **79**, 195327 (2009).

27. T. Lohmann, K. von Klitzing, and J. H. Smet, Nano Lett. **9**, 1973 (2009).

28. J. K. Jain and S. A. Kivelson, Phys. Rev. Lett. **60**, 1542 (1988).

29. J. Martin *et al.*, Nature Phys. **4**, 144 (2008).

30. L. P. Kouwenhoven *et al.*, in *Mesoscopic Electron Transport*, edited by L. L. Sohn, L. P. Kouwenhoven and G. Schon, Kluwer (1997).

**Fig. 1.** (a-b). Schematic and SEM image of a device. (c) $G(V_{bg})$ of a device before (blue) and after annealing (red). (d). Upper panel: $G(V_{bg})$ at $B$=8T. Lower panel: $G$ (color) *vs.* $V_{bg}$ and $B$.

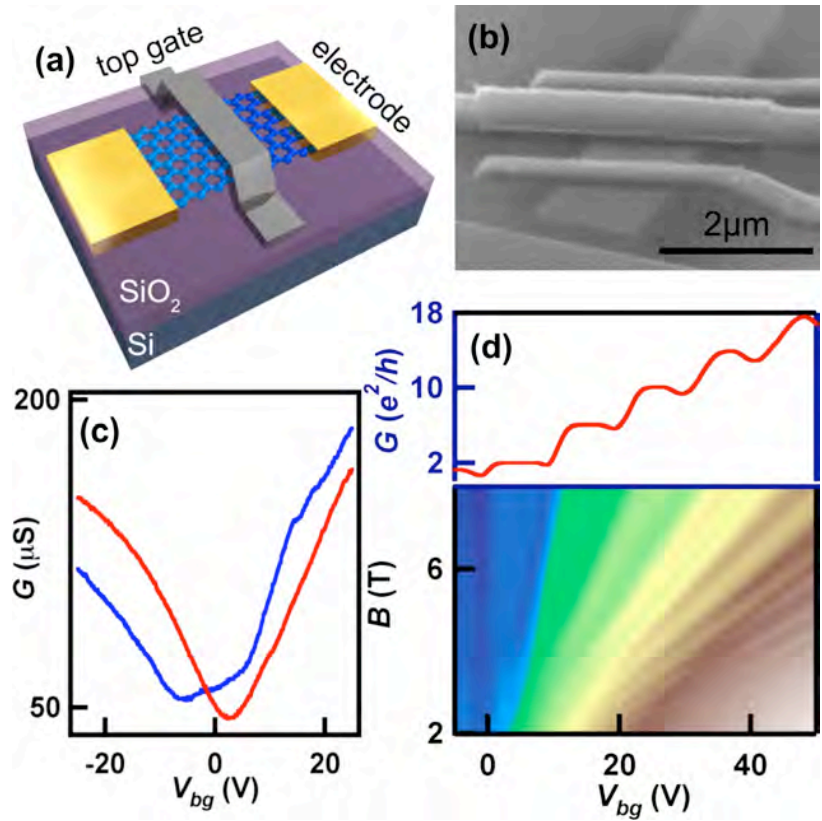

**Fig. 2.** (a) $G(V_{bg}, V_{tg})$ at $B$=8T and $T$=260mK. The curves in the lower panel are line traces along the dotted lines in the color plot, *i.e.* at $v_l$=2, 6 and 10, respectively.

(b). Upper panel: same as (a) but at high gate voltage resolutions. Middle panel: $dG/dV_{tg}(V_{bg}, V_{tg})$ by differentiating data in the upper panel. Lower panel: (red curve) line trace along the dotted line in the upper panel; (blue curve) same data as the red curve but taken with medium $V_{tg}$-resolution. It is offset for clarify.

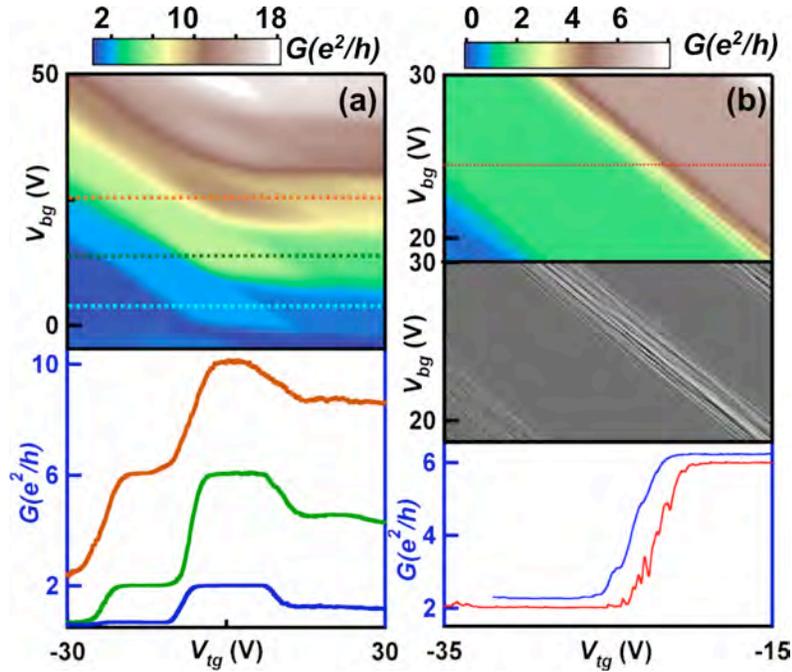

**Fig. 3.** $dG/dV_{tg}(V_{tg}, B)$ at (a) $V_{tg}$=2.5, or $v_1 \approx 2$, and (b) $V_{tg}$=14 or $v_1 \approx 6$. The dotted lines indicate the trajectory of the QH plateaus in the locally gated area.

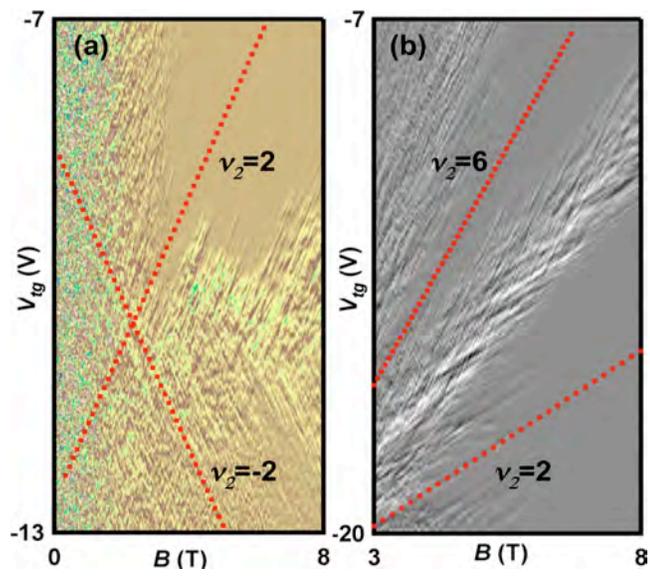

**Fig. 4.** (a). $G(V_{tg})$ for a transition between the $v_2=2$ and $v_2=6$ plateaus. From right to left, $T$=0.26, 0.31, 0.42, 0.53, 0.76, 0.95, 1.43, 1.61, 1.72, 2.35 and 3.8 K. The traces are offset for clarity.

(b). $T$-dependence of $G_p$ (red squares) and $G_d$ (blue dots). The green line is a fit using Eq. (3).

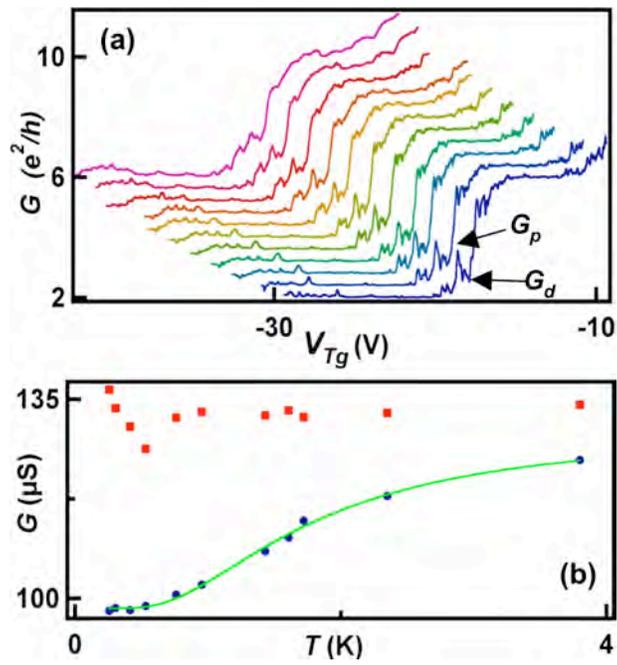